\documentclass[preprint,review]{elsarticle}
\usepackage{amsmath}
\DeclareMathOperator{\sinc}{sinc} \DeclareMathOperator{\tsinc}{tsinc}
\usepackage{lineno,hyperref}
\usepackage[dvipsnames]{xcolor}
\usepackage{subcaption}
\usepackage[utf8x]{inputenc}
\usepackage{chngcntr}
\counterwithout{figure}{subsection}
\journal{Nuclear Inst. and Methods in Physics Research, A}

\makeatletter
\def\ps@pprintTitle{
 \def\@oddfoot{}%
 \let\@evenfoot\@oddfoot}
\makeatother

\begin{document}


\begin{frontmatter}

\title{Positron Annihilation Lifetime Spectroscopy Using Fast Scintillators and Digital Electronics}


\author[mymainaddress]{M. Fang\corref{mycorrespondingauthor}}
\cortext[mycorrespondingauthor]{Corresponding author. Tel.: +1 217 305 1769. Fax.: +1 217 333 2906}
\ead{mingf2@illinois.edu}

\author[mymainaddress]{N. Bartholomew}

\author[mymainaddress]{A. Di Fulvio}

\address[mymainaddress]{Department of Nuclear, Plasma, and Radiological
                        Engineering, \\University of Illinois, Urbana-Champaign,
                        \\ 104 South Wright Street, Urbana, IL 61801, United
                        States}

\begin{abstract}
    Positron Annihilation Lifetime Spectroscopy (PALS) is a non-destructive
    radiological technique widely used in material science studies. PALS
    typically relies on an analog coincidence measurement setup and allows the
    estimate of the positron lifetime in a material sample under investigation.
    The positronium trapping at vacancies in the material results in an
    increased lifetime. In this work, we have developed and optimized a PALS
    experimental setup using organic scintillators, fast digitizers, and
    advanced pulse processing algorithms. We tested three pairs of different
    organic scintillation detectors: EJ-309 liquid, EJ-276 newly developed
    plastic, and BC-418 plastic, and optimized the data processing parameters
    for each pair separately. Our high-throughput data analysis method is based
    on single-pulse interpolation and a constant fraction discrimination (CFD)
    algorithm. The setup based on the BC-418 detector achieved the best time
    resolution of $198.3 ± 0.8$ ps. We used such optimized setup to analyze two
    single-crystal quartz samples and found lifetimes of $156 ± 9$ ps and $366 ±
    22$ ps, in good agreement with the characteristic time constants of this
    material. The proposed experimental set up achieve an excellent time
    resolution, which makes it possible to accurately characterize material
    vacancies by discriminating between the lifetimes of either the spin singlet
    or triplet states of positronium. The optimized data processing algorithms
    are relevant to all the applications where fast timing is important, such as
    nuclear medicine and radiation imaging. 
\end{abstract}

\begin{keyword}
Positron lifetime, organic scintillator, digitizer, CFD 
\end{keyword}

\end{frontmatter}


\section{Introduction}
    PALS is a well-established non-destructive technique used to study defects
    and vacancies in a variety of different materials. In a positron
    annihilation experiment, a positron generating source, such as ${}^{22}$Na,
    is typically placed between two identical samples of a material under
    investigation. ${}^{22}$Na decays into ${}^{22}$Ne through ${\beta}^{+}$
    decay process, creating a positron and an electron neutrino.
    ${}^{22}$Ne then de-excites to its ground state in 3 ps and emits a 1.27 MeV
    gamma ray. The detection of the 1.27 MeV gamma ray can be used to probe the
    creation of the positron. The positrons quickly thermalize through
    scattering and may bind with electrons in the material and form two types of
    positronium: para-positronium (p-Ps) with spin 0 and ortho-positronium
    (o-Ps) with spin 1.The p-Ps decays by emitting two 511 keV annihilation
    photons, while the o-Ps emits three photons in vacuum, as constrained by the
    conservation of angular momentum. In material lattice, the o-Ps mainly
    decays via "pick-off" process where the positron annihilates with a electron
    with opposed spin in the surrounding material and two 511 keV annihilation
    photons are created \cite{brandt1960positronium}. The elapsed time between
    the initial production of the positron and the detection of the annihilation
    photon is therefore a measurement of positronium lifetime in the material
    under investigation.

    The positronium lifetime depends on the material structure. In vacuum, the
    lifetimes of p-Ps and o-Ps are 125 ps and 142 ns, respectively
    \cite{gidley1982new}. The p-Ps lifetime can be affected by the material
    because the Coulomb interaction between the positronium and material changes
    the distance between the positron and electron \cite{saito2003direct}. The
    o-Ps lifetime in a material is reduced drastically due to the "pick-off"
    process. If the material contains voids, vacancies or dislocations, the o-Ps
    can be trapped and the lifetime will be increased compared to the lifetime
    in a defect-free material. Thus, the positronium can be used as a probe to
    investigate the material properties, such as defect density in metals
    \cite{kansy2011pals} and pore characteristics in porous materials
    \cite{gidley1999positronium}. We may also use the PALS to differentiate
    between different lattice structures of the same material since the positron
    lifetime depends on the interaction between the positronium and lattice
    \cite{van2016asymmetric}.

    Time resolution of the measurement system is crucial to perform an accurate
    measurement of positronium lifetime. Hodges and colleagues
    \cite{hodges1972umklapp} set up a system with 330-ps time resolution but
    they were unable to resolve the p-Ps component from the spectra. Haruo and
    Toshio \cite{saito2003direct} achieved a 160 ps time resolution with four
    BaF${}_{2}$ scintillators. However, the slowest component of BaF${}_{2}$
    scintillation light pulse has a decay time of approximately 600 ns, which
    may cause timing artifacts due to pile-up at high count rates. In this work
    we compared the timing performance of three different materials and chose
    the fastest one to perform PALS measurement of single-crystal quartz.
    In recent years, digital electronics, such as digital oscilloscope
    \cite{rytsola2002digital} and fast digitizer \cite{bevcvavr2008high}, are
    replacing traditional analog timing modules in PALS experiments. Digital
    signal processing therefore becomes another important factor affecting the
    time resolution apart from scintillator properties. We have developed a timing
    algorithm based on pulse interpolation and optimized the processing
    parameters for three different organic scintillation detectors. 

\section{Methods}

We measured the time resolution of three different pairs of detectors and
selected the pair that exhibited the best time resolution to then perform the
PALS experiment. We performed a PALS measurement using a ${}^{22}$Na source and
measured the time distribution of the differences of arrival times between the
1.27 MeV ${}^{22}$Na decay gamma ray and the 511 keV annihilation gamma ray.
\subsection{Time Resolution Measurement}
\begin{figure}[ht]
    \centering
    \includegraphics[width=0.8\textwidth]{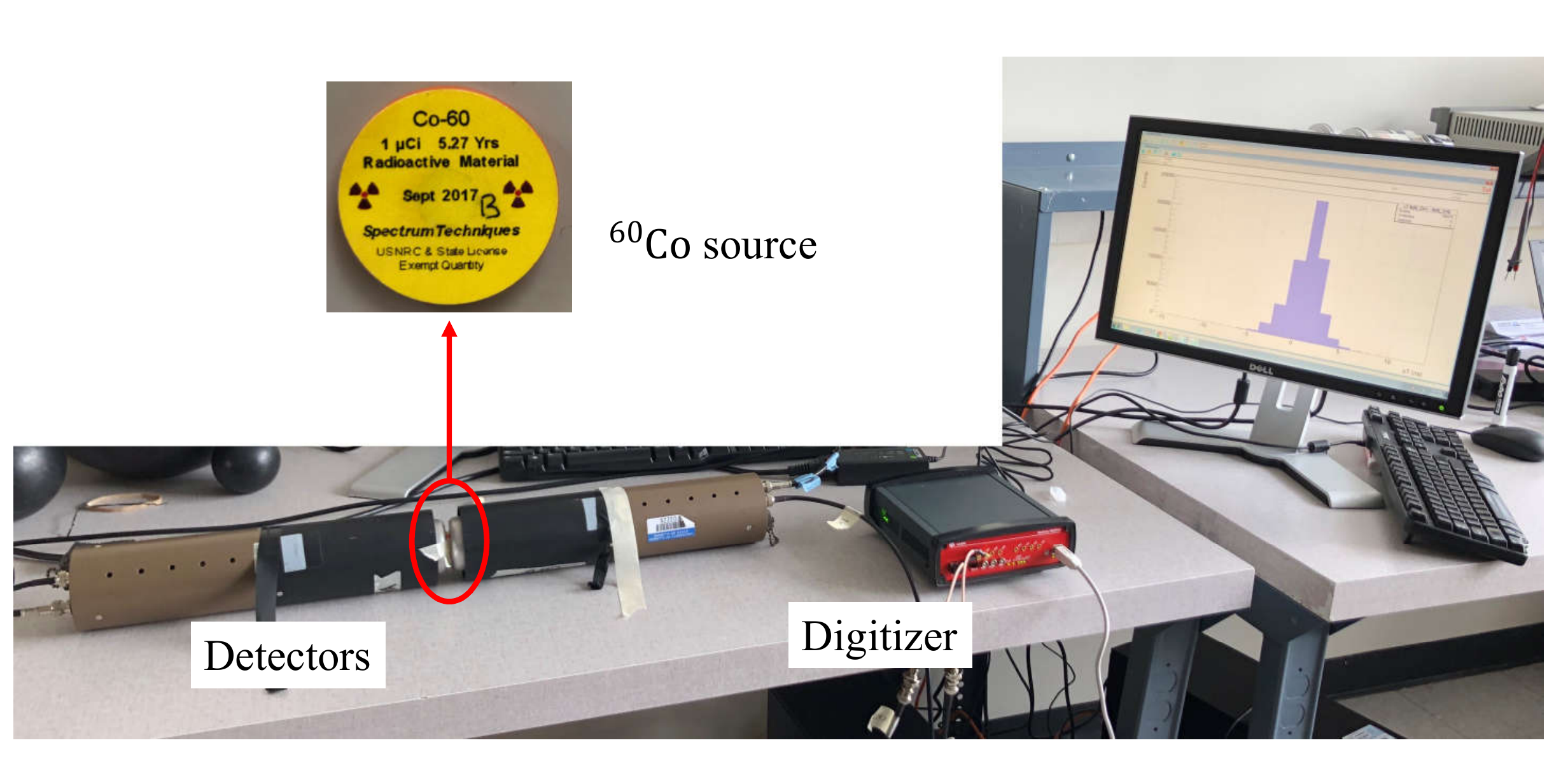}
    \caption{Detector time resolution measurement setup.}
    \label{fig:reso measure}
\end{figure}

We used the experimental setup shown in Fig. \ref{fig:reso measure} for timing
resolution measurement. We performed three measurements with two plastic BC-418,
two liquid EJ-309, and two plastic EJ-276 detectors. Table \ref{table: detector
properties} shows the properties of these detectors.
\begin{table}[ht]
    \caption{Properties of BC-418, EJ-309 and EJ-276 detectors}
    \label{table: detector properties}
    \centering
    \resizebox{\textwidth}{!}{
        \begin{tabular}{cccccccc}
            \hline
            \hline
            Detector & Ratio H:C & Base (cm) & Top (cm) & Height (cm) & Density
            (g/cm${}^3$) & Photomultiplier tube & Pulse shape discrimination\\
            \hline
            BC-418 & 1.100 & 3.18 & 1.27 & 1.27 & 1.032 & R329-02 by Hamamatsu
            Photonics & Not capable \\
            EJ-309 & 1.248 & 5.08 & 5.08 & 5.08 & 0.959 & 9214B by Electron
            Tubes & Capable \\
            EJ-276 & 0.927 & 5.08 & 5.08 & 5.08 & 1.096 & 9214B by Electron
            Tubes & Capable \\
            \hline
            \hline
        \end{tabular}}
\end{table}
    
The time resolution of each detector pair was estimated as the
full-width-at-half-maximum (FWHM) of the distribution of arrival times of two
events occurring in coincidence. In this case, the 1.17 MeV and 1.33 MeV gamma
rays emitted in cascade by a ${}^{60}$Co source are used as reference. A 1µCi
${}^{60}$Co disk source was placed between the two detectors under investigation
in a sandwich configuration Fig. \ref{fig:reso measure}. Approximately 500k
counts in coincidence were collected during each measurement. Detected pulses
were digitized by the 14-bit 500 MS/s digitizer DT5730 by CAEN Technologies and
acquired as full waveforms using the acquisition software CoMPASS
\cite{caencompass} with a 200-ns coincidence window. The detectors were powered
by the Desktop HV Power Supply Module DT5533EN by CAEN Technologies.

We applied the timing algorithm described in Section \ref{algorithms} and
performed a Gaussian fitting of the time difference distribution to obtain the
standard deviation $\sigma$ of the distribution of arrival times and the full
width at half maximum (FWHM = 2.355 $\sigma$).

\subsection{Timing Algorithm}\label{algorithms}
First, we interpolated the digitized pulses. Digitized pulses were acquired by
sampling the analog signal and information about the rising edge and true peak
may be partially lost due to insufficient sampling rates. The sampled signal can
be reconstructed by convolving the samples with the \textit{sinc} function if
the Nyquist condition is satisfied \cite{shannon1998communication}.
\begin{equation}
    g(t) = \sinc (t) * g_s(t) = \sum_{i=-\infty}^{+\infty}g_s(i) \sinc(\frac{t-i \Delta T}{\Delta T}) \label{eq:1}
\end{equation}
Here $g_s(i)$ is the $i$-th sample, $f=1/\Delta T$ is the sampling rate, and the
normalized \textit{sinc} function is defined as
\begin{equation}
    \sinc(x) = \frac{\sin(\pi x)}{\pi x}
\end{equation}
If the time interval between sample $j$ and sample $j+1$ is divided into $N$
even parts, then the $k$-th interpolated value between them is given by
\begin{equation}
    g(j, k) = \sum_{i=-\infty}^{+\infty}g_s(j-i)\sinc(i+k/N) \label{eq:3}
\end{equation}
However, Eq. (\ref{eq:3}) is not suitable for practical use since the sum
extends to infinity and a terminated \textit{sinc} function is used as the
convolution kernel \cite{warburton2017new}, as in Eq. (\ref{eq:4})
\begin{equation} \label{eq:4}
    g(j, k) = \sum_{i=-\infty}^{+\infty}g_s(j-i)\tsinc(iN+k)
\end{equation}
\begin{equation}
    \tsinc(i) = \sinc(i/N)\exp(-(i/T)^2)
\end{equation}
Here \textit{T} is a constant and the Gaussian term quickly drops to 0 as $i$
increases. Thus, the terms in Eq. (\ref{eq:4}) for sufficiently large values of
$i$ can be safely ignored and Eq. (\ref{eq:4}) reduces to a finite sum
\cite{warburton2017new}
\begin{equation}
    g(j, k) = \sum_{i=0}^{L-1}g_s(j-i)\tsinc(iN+k)+g_s(j+1+i)\tsinc((i+1)N-k) \label{eq:5}
\end{equation}
where $L$ is the width of interpolation window.

Afterwards, we applied a digital version of the constant-fraction discrimination
 (CFD) algorithm \cite{steinberger2019timing} to each interpolated pulse and
 obtained the zero-crossing bipolar CFD(i) signal.
\begin{equation}
    \mathrm{CFD}(i) = F\times S(i) - S(i-\Delta) \label{eq:6}
\end{equation}
In Eq. (\ref{eq:6}), $S(i)$ is the value of interpolated pulse at index $i$, $F$
and $\mathrm{\Delta}$ are two constants. $F$ is between 0 and 1, and
$\mathrm{\Delta}$ is a delay time, which is usually comparable to the pulse rise
time. They will be determined later by optimizing the detector time resolution.
The zero-crossing point of the bipolar pulse is defined as the time stamp.

We have implemented the above-mentioned algorithms in a ROOT-based
pulse-processing program\footnote{https://github.com/fm140905/coincidence.git}.
This software allows us to process 1E6 pulses in approximately 10 seconds using
Intel Core i9-7920X @ 2.90GHz.

\subsection{PALS Measurement}
\begin{figure}[ht]
    \centering
    \includegraphics[width=0.6\textwidth]{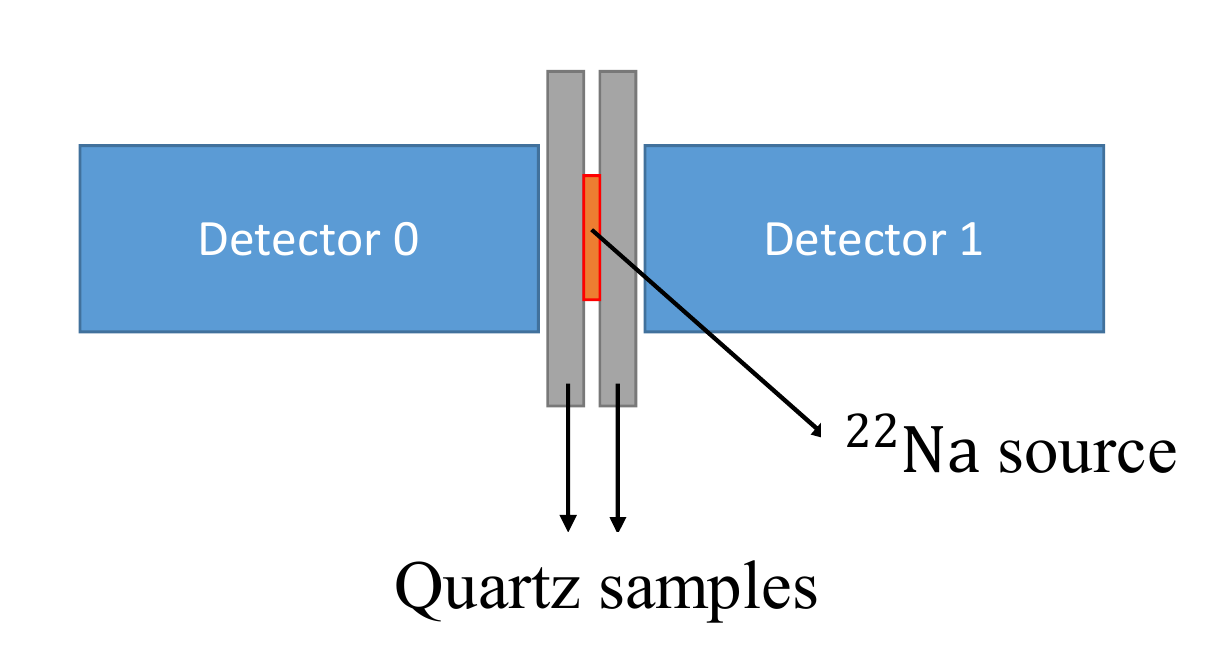}
    \caption{Schematic diagram of the PALS measurement setup using two BC-418 detectors (not to scale).}
    \label{fig:PALS measure}
\end{figure}

The PALS experimental setup is shown in Fig. \ref{fig:PALS measure}. A 10µCi
(1-July-2004) ${}^{22}$Na source was placed in sandwich geometry between two
identical single-crystal quartz samples (10mm × 10mm × 1mm each). ${}^{22}$Na
was sealed between two identical Kapton foils. The quartz samples were purchased
from MTI Corporation. The two scintillation detectors were placed back to back
to achieve the highest detection efficiency. By proper energy gating, detector 0
and detector 1 detected the 1.27 MeV and 511 keV gamma-rays, respectively. We
acquired pulses in coincidence, within a 200-ns time window for 12 hours.

We applied the timing algorithm to pulses and plotted the positron lifetime
spectra. The PALS spectra are usually resolved into three components. The first
component results from the decay of p-Ps, the second one from the mixture of
decays of o-Ps and free positron, and the third one is due to the delayed decay
of o-Ps trapped in defects \cite{van2016asymmetric}. The PALS spectrum can
therefore be modeled as the convolution of exponential decay function and
detector time resolution function, as shown in Eq. (\ref{eq:7}),
\begin{equation}
    f(t) = \sum_{i=1}^3 \frac{I_i}{\tau_i} \mathrm{e}^{-\frac{t}{\tau_i}} * \frac{\mathrm{e}^{-\frac{t^2}{2\sigma^2}}}{\sqrt{2\pi\sigma^2}} 
    =  \sum_{i=1}^3 \frac{I_i}{2\tau_i} \mathrm{e}^{\frac{\sigma^2-2\tau_it}{2\tau_{i}^{2}}} \mathrm{erfc}(\frac{\sigma^2-\tau_it}{\sqrt{2}\sigma\tau_{i}})
    \label{eq:7}
\end{equation}
where $\tau_{i}$ and $I_{i}$ are the lifetime and intensity of the $i$-th
component, $\sigma$ represents the detector time resolution, i.e., FWHM/2.355.

\section{Results}
\subsection{Detector Timing Resolution}
\begin{figure}[ht]
    \centering
    \includegraphics[width=0.6\textwidth]{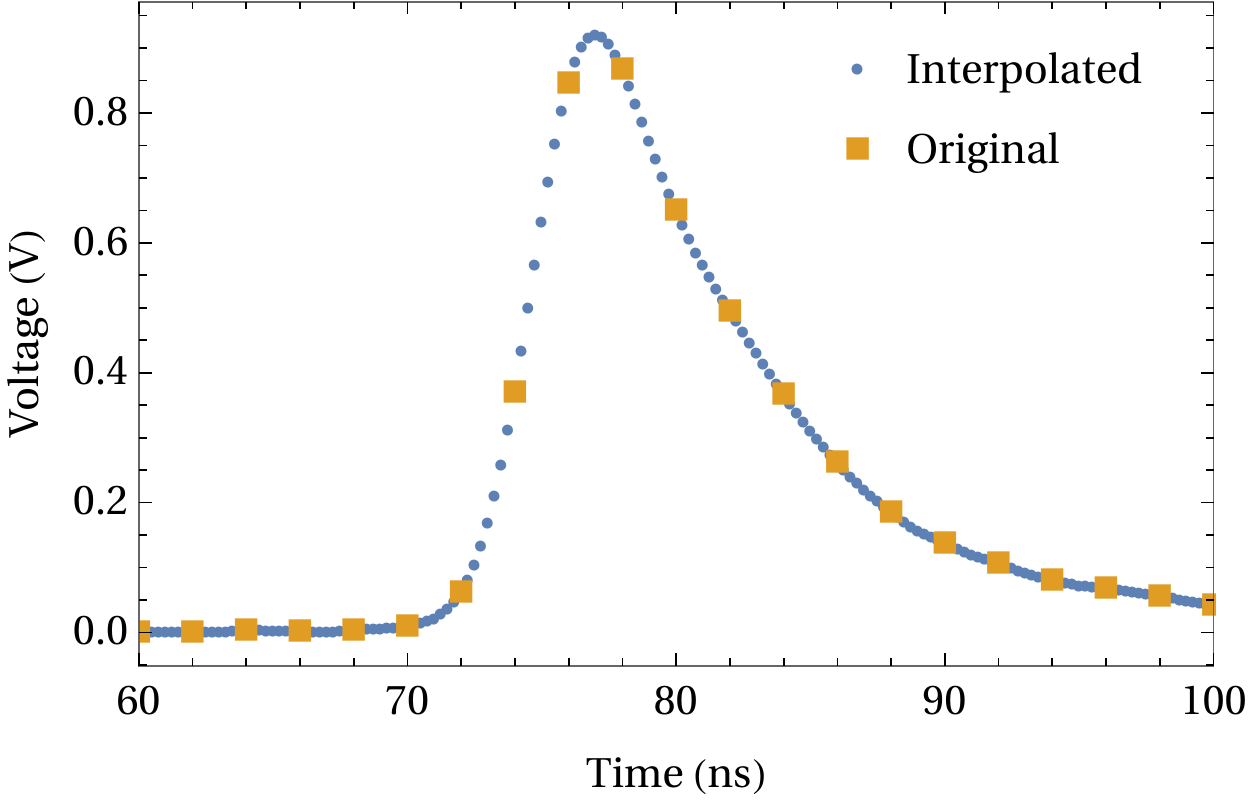}
    \caption{An example pulse before and after interpolation.}
    \label{fig:Example pulse}
\end{figure}

Fig. \ref{fig:Example pulse} shows the comparison between a interpolated pulse and the
original one. The true peak of the original pulse is not captured and only a few
sampling points are recorded on the rising edge due to insufficient sampling
frequency. Interpolation helps in the characterization of the rising edge by
adding more sampling points and gives a more accurate estimate of the true peak.
Since CFD relies on the identification of the time corresponding to the maximum
value and a fraction of it, the time stamp would be more accurate if we perform
CFD after interpolation. As a result, the time resolution would be improved
since it is the spread of the arrival times.

\begin{figure}[ht]
    \centering
    \begin{subfigure}{.5\textwidth}
        \centering
        \includegraphics[width=.9\linewidth]{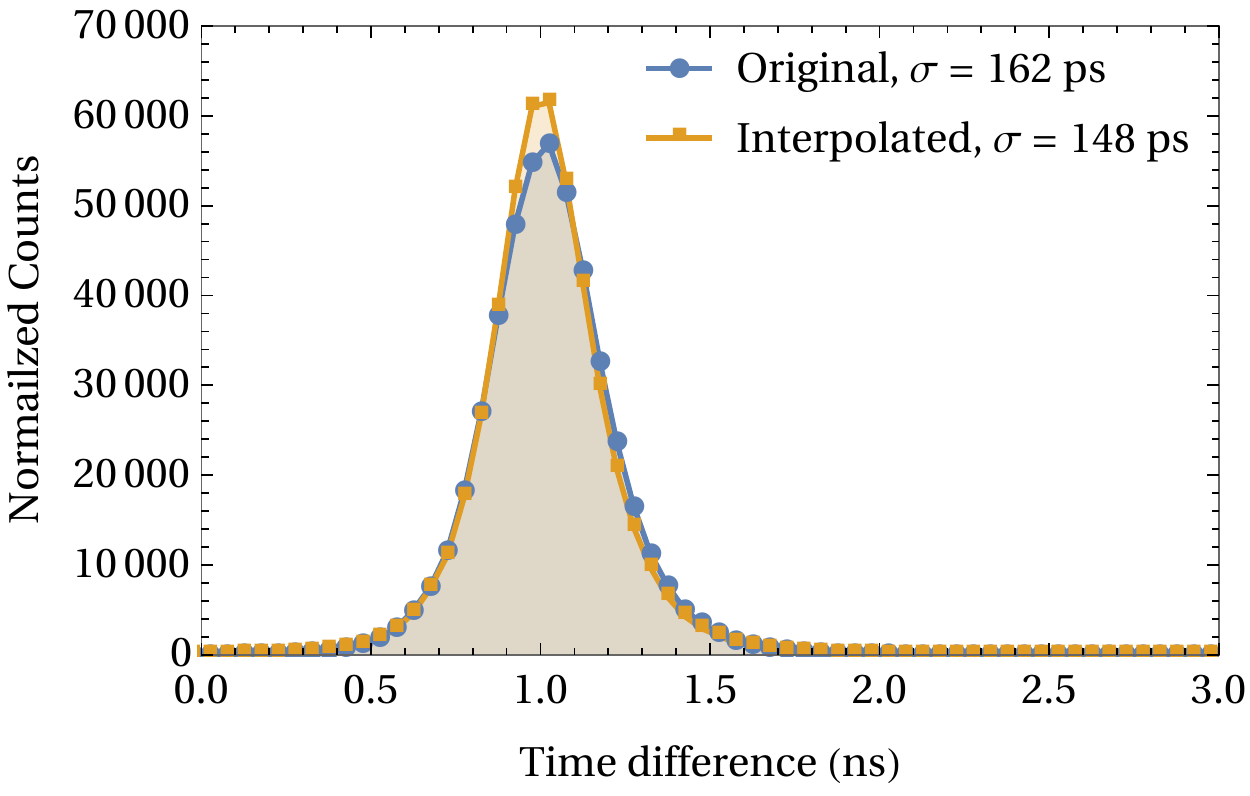}
        \caption{F = 0.75, $\Delta$ = 6 ns}
        \label{fig:6ns}
    \end{subfigure}%
    \begin{subfigure}{.5\textwidth}
        \centering
        \includegraphics[width=.9\linewidth]{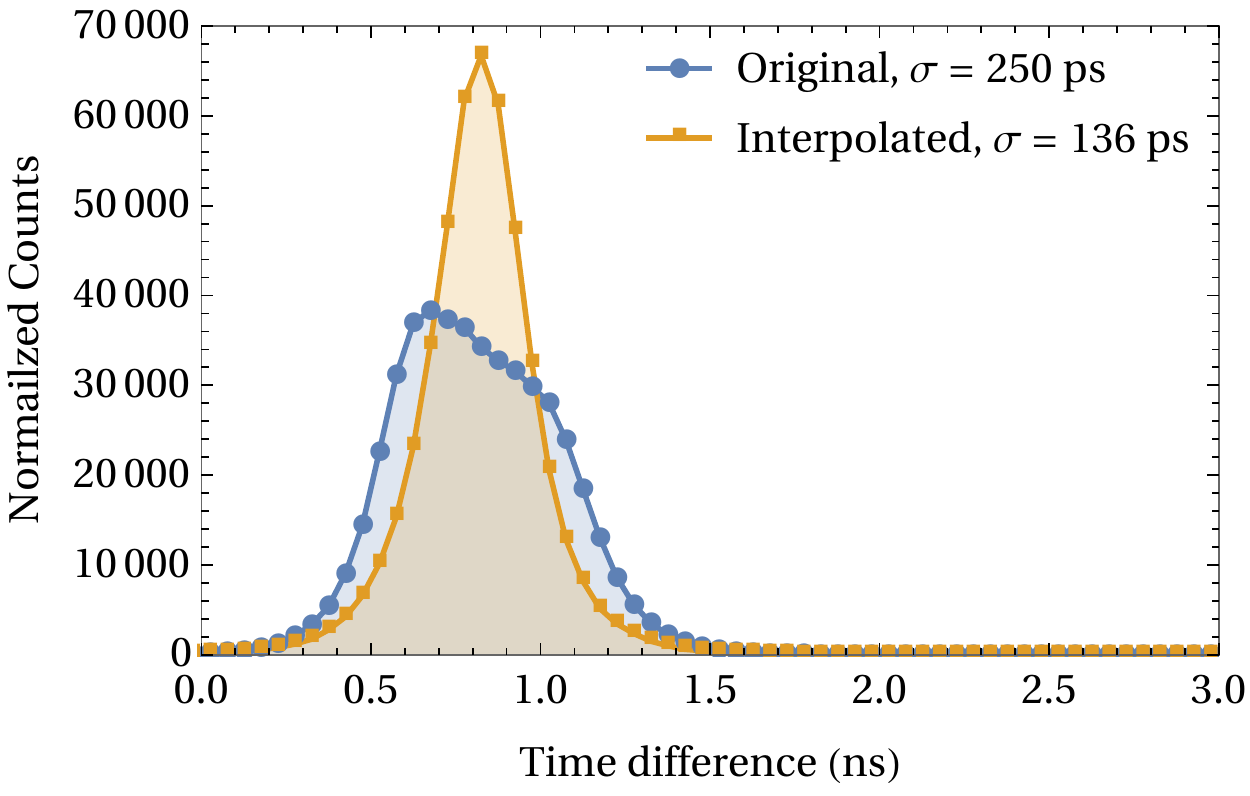}
        \caption{F = 0.4, $\Delta$ = 6 ns}
        \label{fig:4ns}
    \end{subfigure}
    \caption{The time difference distribution before interpolation and after
             interpolation. Measured with BC-418 detectors.}
    \label{fig:BC418_interpolation_comparison}
\end{figure}

Fig. \ref{fig:BC418_interpolation_comparison} shows the time difference
distribution before and after interpolation measured with BC-418 detectors. In
Fig. \ref{fig:6ns}, we performed Gaussian fitting of the spectra to calculate
the FWHM and found that interpolation improved the time resolution by
approximately 33 ps. The spectra are not centered at 0 because of the inherent
asymmetry of acquisition stages, such as slightly different cable lengths.

The interpolation algorithm also helps reduce the skewness of the time
difference histogram. With $\Delta$ fixed, an over-small F usually leads to a
skewed histogram. Fig. \ref{fig:4ns} shows two time difference distributions
before and after interpolation, with F = 0.4 and $\Delta$ = 6 ns. After
interpolation, the time difference histogram is more symmetrical.

\begin{figure}[ht]
    \centering
    \begin{subfigure}{.49\textwidth}
        \centering
        \includegraphics[width=\textwidth]{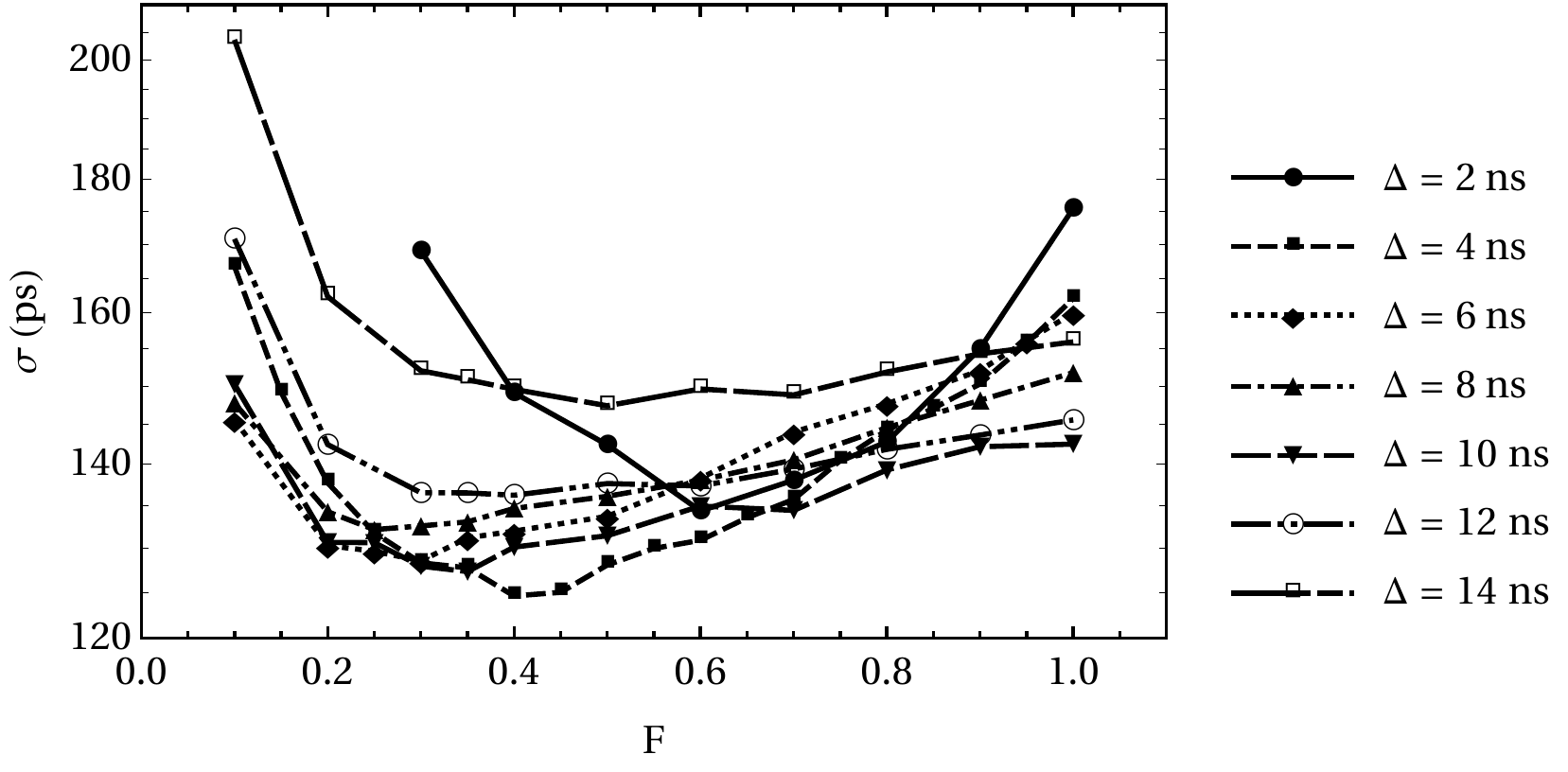}
        \caption{BC-418}
        \label{fig:BC-418 optimization}
    \end{subfigure}%
    \begin{subfigure}{.49\textwidth}
        \centering
        \includegraphics[width=\textwidth]{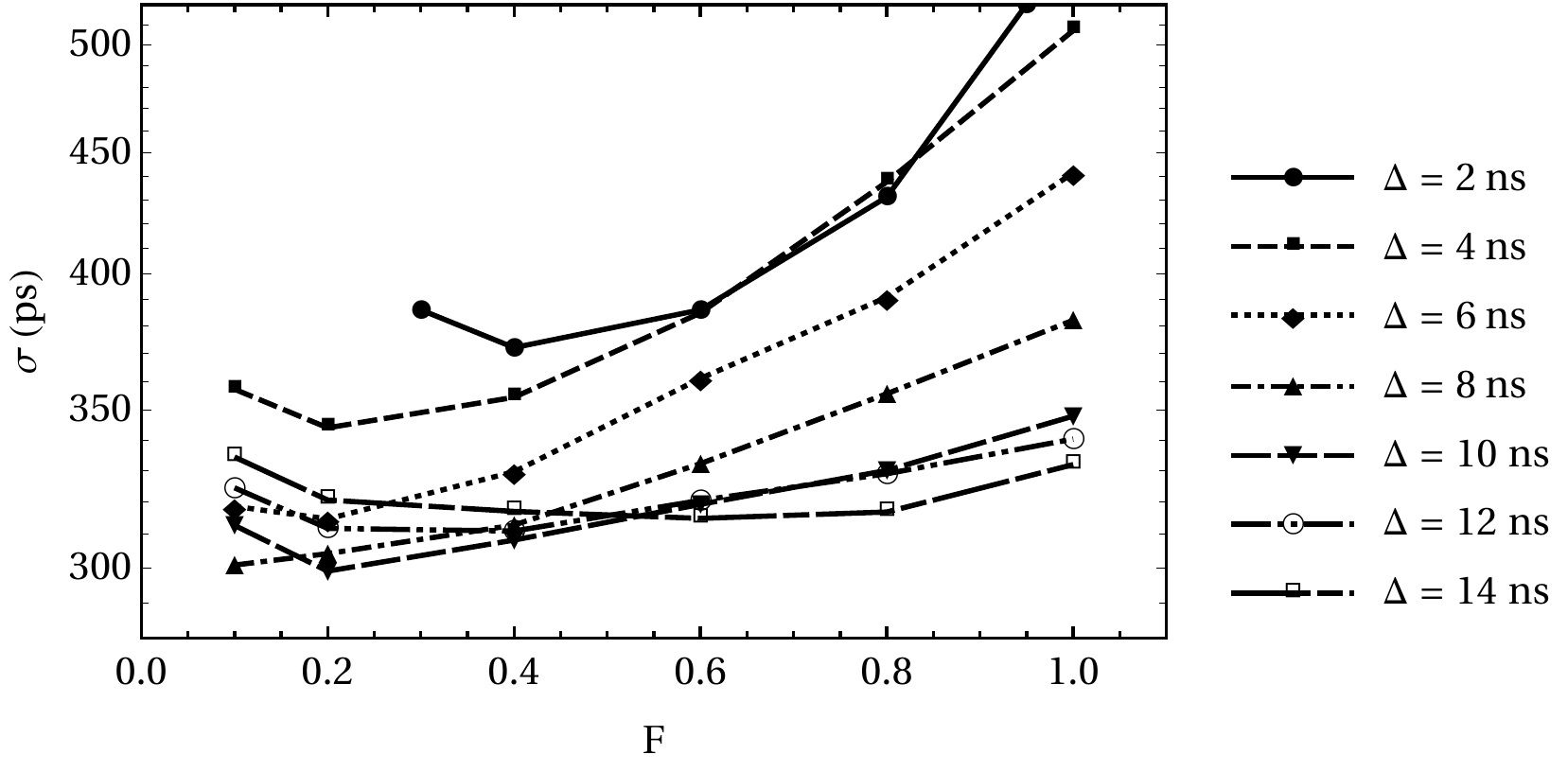}
        \caption{EJ-309}
        \label{fig:EJ-309 optimization}
    \end{subfigure}

    \begin{subfigure}{.49\textwidth}
        \centering
        \includegraphics[width=\textwidth]{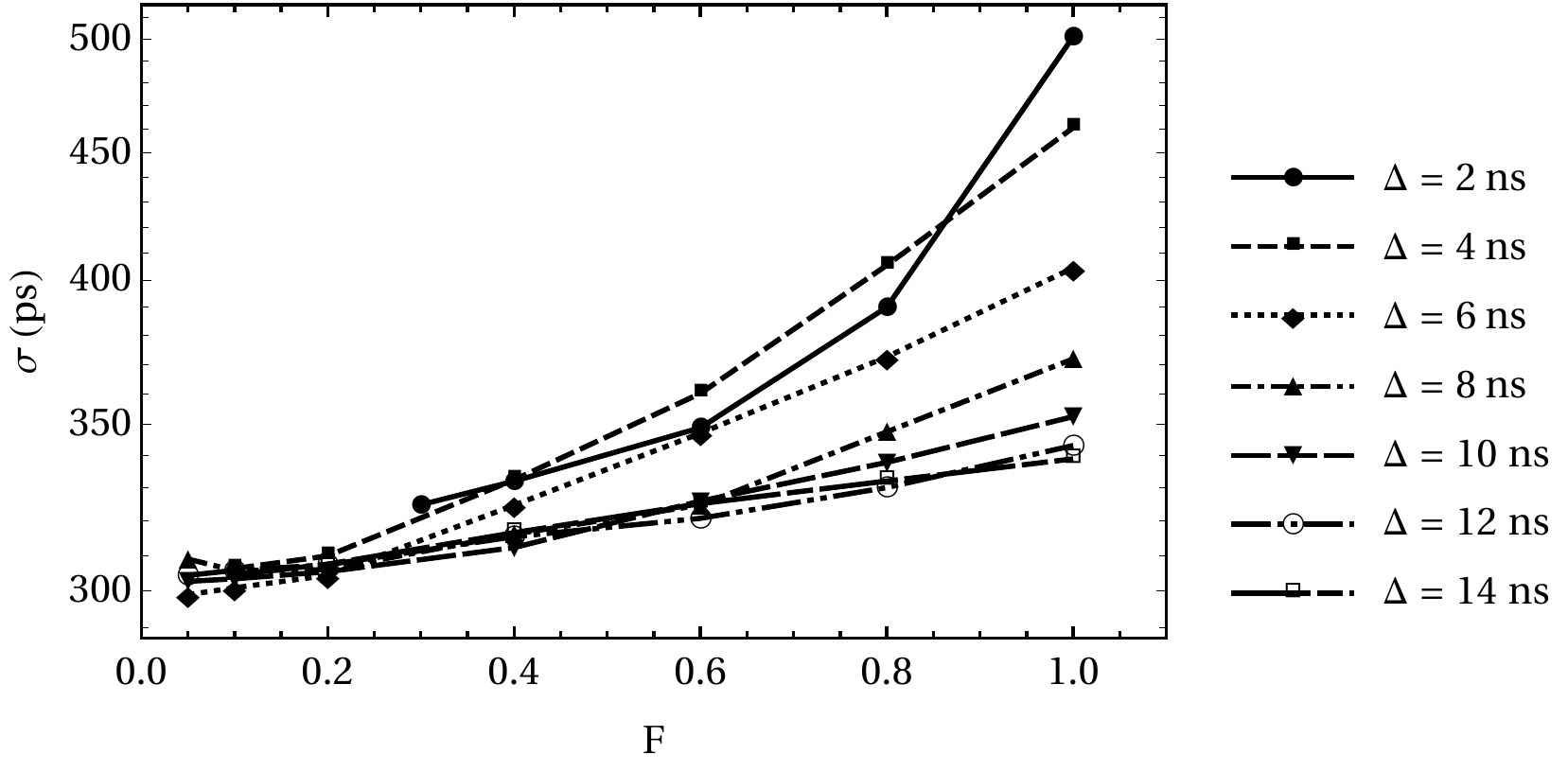}
        \caption{EJ-276}
        \label{fig:EJ-276 optimization}
    \end{subfigure}
\end{figure}

\begin{figure}[ht]
    \centering
    \includegraphics[width=.49\textwidth]{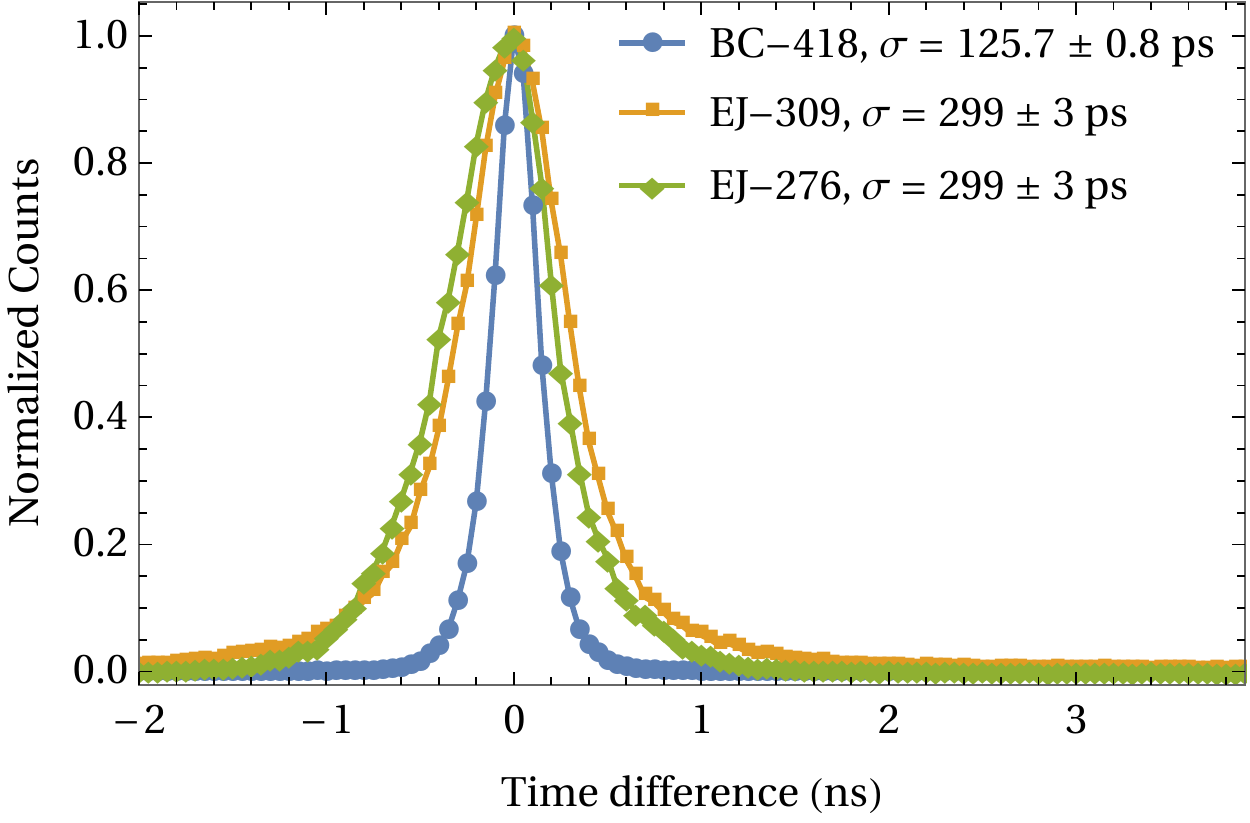}
    \caption{Optimized time resolution}
    \label{fig:optimization compare}
\end{figure}

The FWHM of the time difference distribution depends on the DIACFD parameters F
and $\Delta$. Fig. \ref{fig:BC-418 optimization}, \ref{fig:EJ-309 optimization},
\ref{fig:EJ-276 optimization} illustrate the optimization of F and $\Delta$ for
each detector pair. We increased $\Delta$ in steps of 2 ns and for each $\Delta$
we decreased F from 1 until severe artifacts showed up on the spectrum. For each
combination of F and $\Delta$ we fitted a Gaussian to the spectrum and calculate
the FWHM. Fig. \ref{fig:optimization compare} shows the best time resolution of
each detector pair. Time resolutions of EJ-276 and EJ-309 are close to each
other and BC-418 exhibits the best time resolution. The minimum 195.7 ps
$\sigma$ (293.4 ps FWHM) is obtained with BC-418 detectors at F = 0.4 and
$\Delta$ = 4 ns.

We can further reduce the FWHM by rejecting the low energy pulses. These pulses
have small amplitudes and the sampling values could be easily affected by the
noise, which leads to large errors of the time stamps and creates a long tail in
Fig. \ref{fig:BC418_interpolation_comparison}. After rejecting pulses with
deposited energy less than 600 keVee, the minimum FWHM and optimized parameters
of each detector pair are summarized in Table \ref{table: optimized time
resolutions}. The BC-418 detectors yielded the best time resolution and were
used in the PALS experiment.
\begin{table}[ht]
    \caption{Comparison of detector time resolution}
    \label{table: optimized time resolutions}
    \centering
    \resizebox{0.6\textwidth}{!}{
        \begin{tabular}{ccccc}
            \hline
            \hline
            Detector & $\Delta$ (ns) & F & $\sigma$ (ps) & FWHM (ps) \\
            \hline
            BC-418 & 4 & 0.4 & 84.2 ± 0.3 & 198.3 ± 0.8 \\
            EJ-309 & 10 & 0.2 & 172.5 ± 0.3 & 406.3 ± 0.8 \\
            EJ-276 & 6 & 0.05 & 215.1 ± 0.3 & 507.4 ± 0.7\\
            \hline
            \hline
        \end{tabular}}
\end{table}
\subsection{Positron Lifetime in Single-Crystal Quartz}
\begin{figure}[ht]
    \centering
    \includegraphics[width=0.5\textwidth]{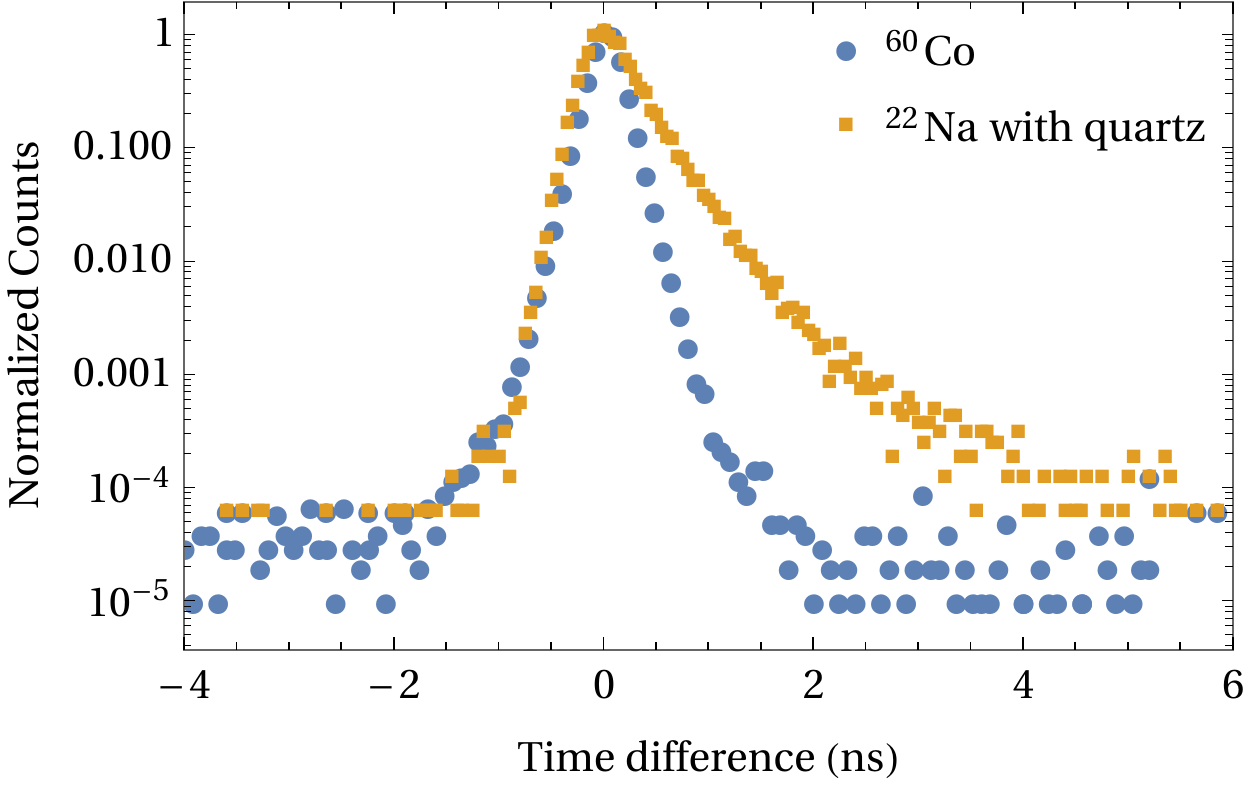}
    \caption{The ${}^{60}$Co spectrum and the positron lifetime spectrum in
            single-crystal quartz, measured with BC-418 detectors.}
    \label{fig:Co60vsNa22}
\end{figure}

Fig. \ref{fig:Co60vsNa22} shows the comparison of the positron lifetime spectrum
in quartz and the distribution of arrival times obtained using the ${}^{60}$Co.
The positron lifetime spectrum shows a longer tail due to longer lifetime, as
expected.

\begin{figure}[ht]
    \centering
    \includegraphics[width=0.5\textwidth]{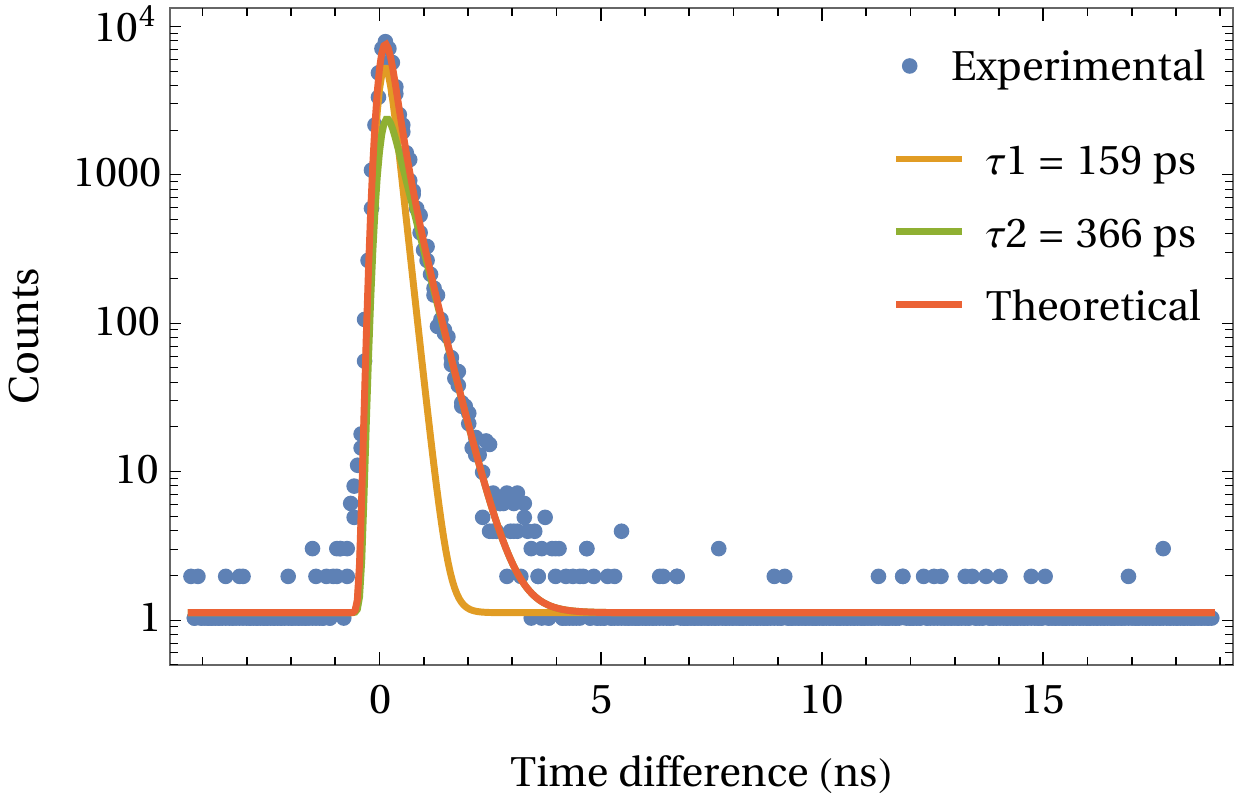}
    \caption{Fit the PAL spectrum of the single-crystal quartz sample.}
    \label{fig:PALSfit}
\end{figure}

Our single-crystal quartz samples are of high purity and high internal
crystalline perfection. Defects such as micro-bubbles and cracks are not allowed
during the manufacturing process. Thus, we believe the third component was
actually undetectable and only two components could be identified from the PALS
spectrum. We fitted Eq. (\ref{eq:5}) to the positron lifetime spectrum using the
LT10 program \cite{kansy2011study}, which is the one of the most widely used
PALS analysis software. The result is shown in Fig. \ref{fig:PALSfit}. The
intensities and lifetimes are shown in Table \ref{table:Fiitted lifetimes}. The
lifetimes $\tau_1$ and $\tau_2$ are 159 ps and 366 ps, respectively, and are in
good agreement with the reported values of 156 ps \cite{saito2003direct} and 358
ps \cite{van2016asymmetric}. $\tau_2$ shows large standard error due to
insufficient amount of counts.
\begin{table}[ht]
    \caption{Lifetimes and intensities}
    \label{table:Fiitted lifetimes}
    \centering
    \resizebox{0.6\textwidth}{!}{
        \begin{tabular}{ccccc}
            \hline
            \hline
            Detector & $\tau_1$ (ps)  & $\tau_2$ (ps) & $I_1$ (\%) & $I_2$ (\%)
            \\
            \hline
            Experiment & 159 ± 9 & 366 ± 22 & 60 ± 6 & 40 ± 6 \\
            Reference & 156 ± 4 & 357 ± 3 & 84.2 ± 0.3 & 15.8 ± 0.3 \\
            \hline
            \hline
        \end{tabular}}
\end{table}

\section{Discussion and Conclusions}
We have implemented a digital version of CFD algorithm to accurately determine
the onset time of interpolated pulses. We also tested another timing algorithm
where we calculated the time when the sampling value exceeds a fixed fraction of
pulse height. The reported method showed better timing resolution. Interpolation
helps in obtaining a non-skewed time difference distribution and improves the
detector timing resolution. There is an optimum value for the F factor, which is
between 0.2 and 0.4 for two of the investigated detectors. BC-418 plastic
detector exhibited the best time resolution, with a FWHM of 198.3 ± 0.8 ps,
because of it's faster response, its truncated-cone geometry and smaller crystal
size. We then used two BC-418 detectors to measure the positron lifetime spectra
in single-crystal quartz and we found the positron lifetimes in quartz were 159
± 9 ps and 366 ± 22 ps. We will use the optimized experimental setup to analyze
vacancies and damages created in radiation detectors irradiated at high fluence
rates. 

\section*{Acknowledgements}
This work was supported by Department of Nuclear, Plasma, and Radiological
Engineering, Grainger College of Engineering and University of Illinois at Urbana-Champaign.


\bibliography{PALS}

\end{document}